\documentclass[twocolumn,showpacs,pre]{revtex4}
\usepackage{amsmath}
\usepackage{graphicx}
\usepackage{dcolumn}
\usepackage{bm}
\usepackage{color}

\begin{document}
\title{Subdiffusion and lateral diffusion coefficient of lipid atoms and molecules
  in phospholipid bilayers}

\author{Elijah Flenner, Jhuma Das, Maikel C. Rheinst\"adter, Ioan Kosztin}

\affiliation{Department of Physics and Astronomy, University of Missouri-Columbia,
  Columbia, MO 65211, U.S.A.}

\date{\today}

\begin{abstract}
  We use a long, all-atom molecular dynamics (MD) simulation combined with
  theoretical modeling to investigate the dynamics of selected lipid atoms and
  lipid molecules in a hydrated diyristoyl-phosphatidylcholine (DMPC) lipid
  bilayer.
  From the analysis of a $0.1~\mu$s MD trajectory we find that the time evolution
  of the mean square displacement, $\langle[\delta{r}(t)]^2\rangle$, of lipid atoms
  and molecules exhibits three well separated dynamical regions: (i) ballistic,
  with $\langle[\delta{r}(t)]^2\rangle \sim t^2$ for $t\lesssim 10~\text{fs}$;
  (ii) subdiffusive, with $\langle[\delta{r}(t)]^2 \rangle \sim t^{\beta}$ with
  $\beta<1$, for $10~\text{ps}\lesssim t \lesssim 10~\text{ns}$; and (iii) Fickian
  diffusion, with $\langle[\delta{r}(t)]^2\rangle \sim t$ for $t\gtrsim
  30~\text{ns}$.
  We propose a memory function approach for calculating
  $\langle[\delta{r}(t)]^2\rangle$ over the entire time range extending from the
  ballistic to the Fickian diffusion regimes. The results are in very
  good agreement with the ones from the MD simulations.
  We also examine the implications of the presence of the subdiffusive dynamics of
  lipids on the self-intermediate scattering function and the incoherent dynamics
  structure factor measured in neutron scattering experiments. 
\end{abstract}


\maketitle


\section{Introduction}
\label{intro}

\begin{figure}
\includegraphics[width=3.2in]{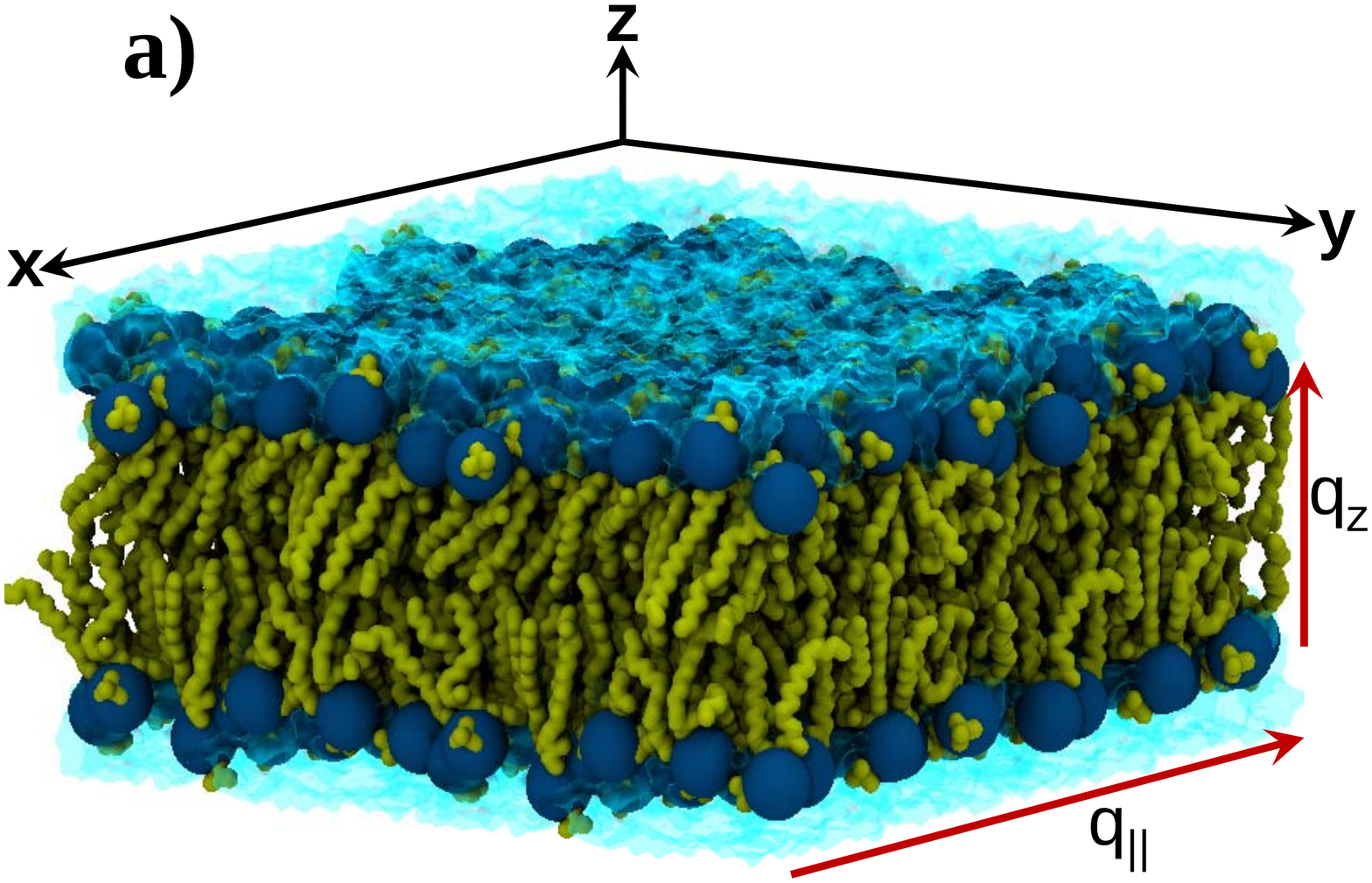}
\includegraphics[width=3.2in]{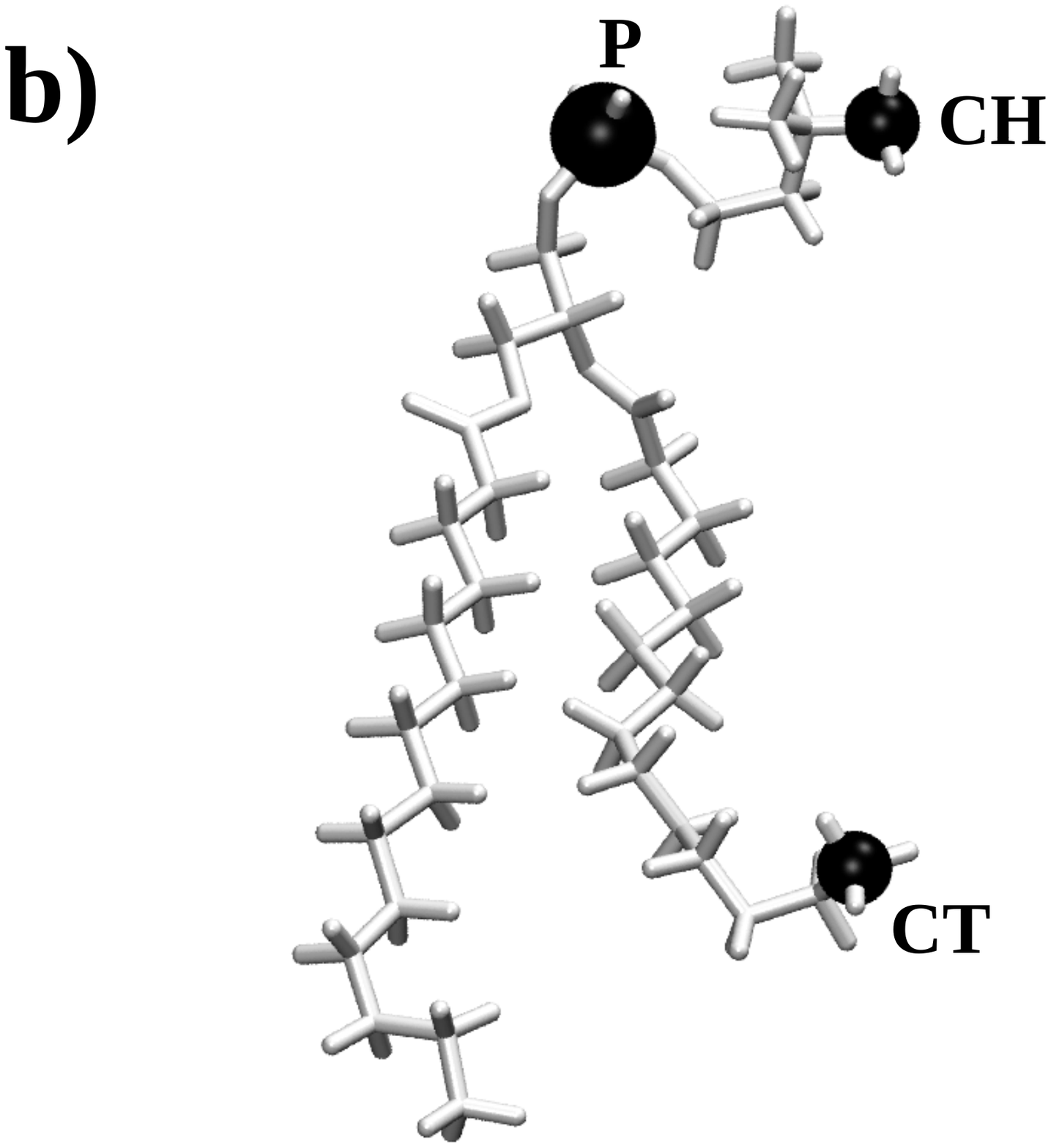}
\caption{(color online) (a) The simulated DMPC lipid bilayer system. (b) A DMPC
  lipid molecule. The atoms whose dynamics is investigated in this work (P, CH and
  CT) are highlighted as black spheres. Figures created using VMD \cite{VMD}. 
\label{lipid}}
\end{figure}

Lipids are polymer molecules composed of hydrophobic acyl chains attached to a
hydrophilic polar head group.  In the presence of a polar solvent (e.g., water)
lipids can spontaneously self-assemble to form a bilayer membrane
(Fig.~\ref{lipid}).
The fluid (L$_\alpha$) phase of lipid bilayers behaves as a two dimensional (2D)
fluid and the lateral 2D self-diffusion coefficient of individual lipids within a
leaflet of the bilayer has been determined both experimentally by a variety of
methods \cite{Pfeiffer1988,Tabony1990,Vaz1985,Wennerstrom1977,Kuo1979,Shin1991,Koenig1995,Pfeiffer1989,Koenig1992,Almeida1995}
and through computer simulations
\cite{Falck2008,Wohlert2006,Doxastakis2007,Lindahl2001,Anezo2004,Patra2004,Klauda2006,Hofsass2003,Falck2004}.
The experiments suggests that there are at least two relevant length/time scales
associated with the lateral diffusion of lipids in a bilayer.
Experiments designed to probe motion on picosecond (ps) time scales measure a
diffusion coefficient $D_1$ that can be one to two orders of magnitude larger than
the diffusion coefficient $D_2$ measured in experiments which probe motion on tens
or hundreds of nanosecond (ns) time scales.

There have been several models used to explain the difference in the diffusion
coefficients $D_1$ and $D_2$. Vaz and Almeida \cite{Vaz1991} suggested a
free-volume jump-diffusion model where the lipid moves in discrete steps when a
void forms next to the lipid. After a jump, the lipid can either return to its
original position or another lipid can jump into the empty space left behind. 
In this model, the short time diffusion coefficient $D_1$ is associated with the
``rattling'' motion of the lipid inside the ``cage'' created by the
surrounding lipids.
The jump diffusion mechanism was investigated by Falck \textit{et al.}\
\cite{Falck2008} in a simulation of diyristoyl-phosphatidylcholine
(DMPC) bilayers, and they did not observe enough
jump events to provide evidence for the jump diffusion model.  
To explain their MD simulation results of a DMPC bilayer, Wohlert and Edholm
\cite{Wohlert2006} proposed a model where the motion of the lipid in the plane of
the bilayer can be regarded as diffusion (with a coefficient $D_1$) confined in a
circular ``cage'' whose center undergoes free diffusion (with a coefficient
$D_2$). The two diffusion coefficients obtained by fitting their simulated data to
the analytical solution of this model were in reasonable agreement with the
experimental values. 

None of the above diffusion models explicitly take into account the
microscopic polymeric structure of lipids, but effects of this structure
has been observed in simulations.
For example, Doxastakis \textit{et al.}\ \cite{Doxastakis2007} observed in a MD
simulation of 1,2-Dipalmitoyl-sn-Glycero-3-Phosphocholine
(DPPC) bilayers that the atoms at the ends of the lipid tails fluctuate
more than those in the head group, but the overall diffusion of the lipid is
limited by the diffusion of the head group. This observation led them to examine a
diffusion in a sphere model where the size of the sphere depends on the position
of the chain. However, the authors describe this model as ``not particularly
accurate'', and the model was extended to include a distribution of spheres at each tail position.

The dynamics of the atoms in lipid molecules are more complex than those in
ordinary liquids.
In a simple liquid \cite{Hansen2006} atoms move ballistically at short times.
Thus the mean square displacement (MSD) $\langle [\delta r(t)]^2 \rangle \approx
\langle (v t)^2 \rangle \sim t^2$, which is followed by a crossover to Fickian
diffusion, characterized by $\langle [\delta r(t)]^2 \rangle \sim t$ for long
times. In dense fluids a caging effect, where the atoms are trapped by their
neighbors, is observed between the ballistic and diffusive regimes, leading to a
plateau in $\langle [\delta r(t)]^2 \rangle$.  In a lipid bilayer, the motion of
lipid atoms is further complicated by the polymeric structure, characterized by
connectivity and flexibility, of the lipids.
 
A more realistic description of the diffusion of lipids should be based on
theoretical models designed for polymers.  Two such theories are the Rouse model
\cite{Strobl2007} and the mode-coupling theory \cite{Chong2002,Chong2007}.
In the Rouse model over-damped Brownian motion is assumed for the individual
monomers with a harmonic potential connecting adjacent monomers. 
 The motion is subdiffusive with $\langle [ \delta r(t)]^2 \rangle \sim
t^{1/2}$ then there is a crossover to diffusive motion $\langle [\delta r(t)]^2 \rangle \sim t^{1}$ 
at later times.
%
The mode-coupling theory (MCT), generally used to study the dynamics of
glass-forming liquids, has recently been extended for flexible macromolecules and
polymers \cite{Chong2002,Chong2007}, though it has not been applied to study lipid
bilayers to our knowledge.  The MCT is a method used to
approximately calculate space-time correlation functions (e.g., intermediate
scattering function) by solving a set of integro-differential equations obtained
from first principles.
An important feature of the extended polymer version of the mode coupling theory
is the prediction of a sub-diffusive region between the short time ballistic and
the long time Fickian diffusion regimes.  Specifically, the theory predicts
$\langle [\delta r(t)]^2 \rangle \sim t^{\beta}$, with $\beta$ typically between
0.5 and 0.65. Besides other parameters, the exponent
$\beta$ depends on the length of the polymer chain and approaches 0.5, the Rouse
limit, for an infinite chain of identical monomers\cite{Chong2002,Chong2007}.
Since the only input to the theory is the static structure, 
the subdiffusion predicted by the extended MCT is due to the polymeric
structure and not to the specifics of the interactions. Thus, one expects that the
dynamics of the atoms in a lipid bilayer should exhibit 
a pronounced sub-diffusive regime before it crosses over to normal
Fickian diffusion.
We argue that the extended sub-diffusive region in the MSD is the cause of the
difference between the experimentally measured diffusion coefficients $D_1$ and
$D_2$.

In this paper, we report an extensive computational investigation of the complex
 dynamics of both individual lipid atoms and entire lipid molecules in
the fluid phase of a hydrated DMPC bilayer by
employing a microscopic description of the system. Our study is based on a $
0.1~\mu$s long all atom molecular dynamics (MD) simulation of the DMPC bilayer.
By calculating the time evolution of the mean square displacement of lipid atoms
in the plane of the bilayer for a time interval ranging from $10^{-15}$s to
$10^{-7}$s, we identify the short time ($t < 0.1$~ps) ballistic region, the long
time ($t \gtrsim 30$~ns) Fickian diffusion region, and in between an extended
\emph{subdiffusive} region characterized by a power law $\langle [\delta{r}(t)]^2
\rangle \sim t^{\beta}$, with $\beta<1$.  We find that $\beta$ depends on the atom
type and, most importantly, on its position within the lipid molecule. The
dynamics is different for atoms in the lipid tail versus atoms in the head group.
We also examine the implications of the heterogeneous subdiffusion of the
individual atoms on the diffusion of the lipid molecule as a whole, and on the
interpretation of neutron scattering experiments.

This paper is organized as follows. In Sec.~\ref{sim} we describe
our MD study of a solvated DMPC lipid bilayer. In
Sec.~\ref{msd} we provide a detailed numerical and theoretical study of the 
mean square displacement of lipid
atoms and lipid molecules. In Sec.~\ref{sf} we examine the implications of
this study for the incoherent intermediate scattering function, and the dynamic
structure factor that is measured in neutron scattering experiments. 
Lastly, in Sec.~\ref{sec:conc} we present a discussion and our conclusions.

\section{MD Simulation of DMPC Bilayer}
\label{sim}

To investigate the dynamics of selected lipid atoms and of entire lipid molecules,
we performed a $0.1~\mu$s long all atom MD simulation of a fully
solvated DMPC lipid bilayer (Fig.~\ref{lipid}).  The pre-equilibrated structure of
the bilayer was obtained from Mikko Karttunen's web site
(www.apmaths.uwo.ca/$\sim$mkarttu/downloads.shtml) \cite{Gurtovenko2004} and
contained 128 lipid molecules. The system was solvated by adding two 11~\AA~thick
layers of water to each side of the membrane using the \textit{Solvate} plugin in
VMD \cite{VMD}. The final system contained a total of 2577 TIP3
\cite{Jorgensen1983} water molecules.
The MD simulation was performed with NAMD-2.6 \cite{NAMD} using the CHARMM27
\cite{Feller2000} force field.  The equations of motion were integrated with a
multiple time step algorithm with time steps of 1~fs for bonding interaction, 2~fs
for non-bonding interactions, and 4~fs for long-range electrostatic interactions.
The non-bonded interactions were cutoff at 12~\AA\ with a smooth switching
function starting at 10~\AA.  Long-range electrostatic interactions were computed
by employing the smooth Particle Mesh Ewald method \cite{essmann95-8577}
with a grid spacing of 1~\AA.

The system was brought to $T=303$~K and normal pressure ($p=1$~atm) through
several stages of equilibration.  First, the system was subjected to $6\times
10^4$ energy minimization steps by harmonically restraining the phosphorous (P)
atoms of the lipid headgroups along the normal direction to the surface of the
membrane. Next, the system was gradually heated to the desired temperature
$T=303$~K during a 0.5~ns period. After removing all the restraints, the system
was equilibrated through a 75~ns long NpT simulation. At the end of this process
the area per lipid (APL) was 56.2~\AA$^2$. To increase the APL to the desired
value of $\sim 60$~\AA$^2$, we performed NVT simulations and gradually increased
the size of the system in the $xy$-plane of the membrane. The final system size
was $62\times62\times58.5$~\AA$^3$, with an APL of $60.06$~\AA$^2$. The
equilibration process was concluded with an additional 10~ns NVT simulation.

A production run of $0.1~\mu$s was performed in the NVT ensemble. We employed a
Langevin thermostat with a coupling constant of 0.05~ps$^{-1}$. The coordinates of
all the atoms were stored every 2~fs for the first 100~ps, then saved every 20~fs
for the next 10~ns, then every 100~fs for the next 90 ns. This allowed us to study
the short, intermediate, and long time dynamics of the individual atoms and
lipids. A snapshot of the lipid bilayer system is shown in Fig.~\ref{lipid}a.
The MD simulations were carried out on 40 CPUs of a dual core 2.8GHz Intel Xeon
EM64T cluster with a performance of around 0.2 days/ns.

\section{Mean Square Displacement}
\label{msd}

In this section we examine the time dependence of the lateral mean square
displacement (MSD) of selected atoms within a lipid, and of entire lipids within
the bilayer. We compare the results with the general predictions of the extended
MCT for polymers \cite{Chong2002,Chong2007}, and we propose a memory function
method capable of describing the time dependence of MSD in lipid bilayers.
 
We begin by outlining some of the qualitative predictions of the extended MCT for
polymers.
To derive the extended mode coupling equations, one first applies the Mori-Zwanzig
projection operator formalism in the so called site-site formalism.  This leads to a
set of coupled integro-differential equations for monomer density
fluctuations. The MCT is a set of approximations of the memory kernel that permits the
corresponding integro-differential equations to be numerically solved with the
only input being the static structure.
The monomer and the polymer center of mass MSD can be computed from the solution
of the mode-coupling equations from the small wave-vector, $q$, limit (see
Ref.~\onlinecite{Chong2007} for details).  For the MSD $\langle [\delta r^a(t)]^2
\rangle = \langle [\vec{r}\,^a(t) - \vec{r}\,^a(0)]^2 \rangle$ for monomer $a$,
the MCT predicts an initial ballistic region where $\langle [\delta r^a(t)]^2
\rangle \sim t^2$. After the ballistic regime, the growth of the mean square
displacement is suppressed due to a local caging effect caused by the trapping of
the polymer by its neighbors. At larger times there is a crossover to Fickian
diffusion with $\langle [\delta r^a(t)]^2 \rangle = 2d D_L t$, where $d$ is the
dimension of the system and $D_L$ is the long time self diffusion coefficient for
the monomer.  In this crossover region there is a sub-diffusive region where
$\langle [\delta r^a(t)]^2 \rangle \sim t^\beta$ with $\beta<1$.  The value of
$\beta$ is smaller for longer chains and has a limiting value of 0.5 for an
infinitely long chain.  Note that the sub-diffusive region is absent for simple
liquids, and is a \textit{polymer} specific feature.

\begin{figure}
\includegraphics[width=3.2in]{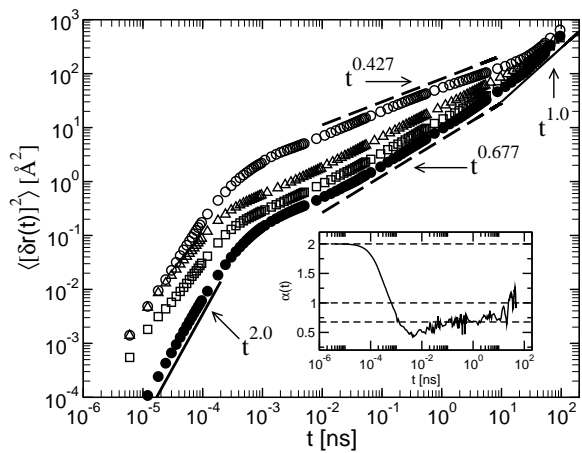}
\caption{Lateral mean square displacement of the P (squares), CH (triangles) and
  CT (open circles) atoms, along with the center of mass (CM) of the lipids
  (closed circles).  The selected atoms (P, CH and CT) are highlighted in
  Fig.~\ref{lipid}. Inset: $\alpha(t) = \partial \ln \langle [\delta r^{CM} (t)]^2
  \rangle  /\partial \ln(t)$. The dashed horizontal
  lines correspond to the ballistic ($\alpha=2$), sub-diffusive ($\alpha=0.677$)
  and Fickian diffusion ($\alpha=1$) values.}
\label{Pmsd}
\end{figure}

To test the MCT predictions for the lipids, we examined the motion of three
representative atoms in the lipid (see Fig.~\ref{lipid}b), namely the phosphorous
(P), the carbon C15 (CH) atoms in the lipid head group, and the carbon atom C214
(CT) located at the end of one of the lipid tails. Because the surroundings of the
head group atoms is different from those in the lipid tails, we expect that the
MSD of the P and CH atoms will show distinctive features from that of the CT
atoms.
Shown in Fig.~\ref{Pmsd} is the time evolution of the lateral MSD of the P atom (squares)
$\langle [\delta r^P(t)]^2 \rangle = \langle |r^P(t)-r^P(0)|^2 \rangle$ on a
log-log scale, which follows the qualitative predictions of the MCT. Specifically,
an initial ballistic region is followed by a narrow caging region that gradually
crossover to linear diffusion through an extended sub-diffusive region
characterized by $\langle[ \delta r^P(t)]^2 \rangle \sim t^{\beta}$, with $\beta <
1$.

Also plotted in Fig.~\ref{Pmsd} is the MSD for the CH (triangle) and CT (circles)
atoms.  The ballistic, the sub-diffusive, and the diffusive time scales are
identifiable for all the atoms. The MSD in the ballistic regime is the same for CH
and CT since they have the same mass and are at the same temperature, but their
mean square displacements differ starting in the caging region. The MSD for CT is
larger than that of CH, which indicates that the size of the cage is larger for
atoms at the end of the tail than atoms close to the head group. This difference
is a direct consequence of the structure of the lipid bilayer, and has been
observed in previous simulations \cite{Doxastakis2007,Wohlert2006}.

For all the atoms the crossover to Fickian diffusion begins around
10~ns and $\langle [\delta r^a(t)]^2 \rangle\approx 100$~\AA$^2$, which 
approximately corresponds to
the nearest neighbor distance of the lipids ($\sim 10$~\AA). Thus the linear
diffusion can be observed only after the lipid has moved around one lipid
diameter.

To determine the power-law exponent $\beta^a$ for atom $a$ in the sub-diffusive
region, we calculated 
\begin{equation}
  \label{eq:alpha}
  \alpha^a(t) = \frac{\partial \ln[\langle [\delta r^a (t)]^2 \rangle
]}{\partial \ln(t)} =
\frac{t}{\langle[\delta{r}^a(t)]^2\rangle}\frac{\mbox{d}\langle[\delta{r}^a(t)]^2\rangle}{\mbox{d}t} ,
\end{equation}
and averaged the result from $t = 0.01$~ns to $t = 10$~ns, the region
where $\alpha^a(t)$ is approximately constant. The result for $\alpha^a(t)$ for the
center of mass (CM) of the lipids is shown in the inset to Fig.~\ref{Pmsd}.  The
horizontal lines in the inset correspond to the ballistic ($\beta=2$),
sub-diffusive ($\beta=0.677$) and Fickian diffusion ($\beta=1$) regimes.
The value of $\beta^a$ depends on the atom type and ranged from $0.677 \pm 0.005$
for the CM to $0.427 \pm 0.004$ for CT, the carbon atom at the end of the
tail. The dynamics in the ps time scale strongly depends on the position of the
atom within the lipid. The atoms at the end of the lipid tails have more freedom
to move, and hence a larger cage. However, these atoms are connected to those
within the head group, which ultimately determine their diffusion on long time
scales.

For $t \gtrsim 30$~ns, $\alpha^a(t) \approx 1$ for each of the individual atoms
($a=$ P, CH, CT) and the center of mass, thus the lipids appear to be undergoing
Fickian diffusion.  We calculated the diffusion coefficient $D^a_L = \lim_{t
  \rightarrow \infty} \langle [\delta r^a(t)]^2 \rangle/4t$ by fitting $\langle
[\delta r^a(t)]^2 \rangle$ for $t>30$ ns to a straight line. The diffusion
coefficients ($D_L$-msd) are shown in Table~\ref{Dtable}, and their values range between
$D^{\text{CT}}_L=1.67 \times 10^{-7}$~cm$^2$s$^{-1}$ and
$D^{\text{CH}}_L=1.27\times 10^{-7}$~cm$^2$s$^{-1}$.
These values are within statistical error and are close to the diffusion
coefficient of the CM. However, these diffusion coefficients are almost a factor
of two larger than the value of $D^{\text{exp}}=0.69 \times
10^{-7}$~cm$^2$s$^{-1}$ obtained from FRAP measurement 
\cite{Almeida1995} of DMPC at 34$^\circ$C. 
This difference may be due to a combination of several factors, e.g., imperfection
of the empirical CHARMM27 force field used in the MD simulations, discrepancy
between the APL of the experimental and the simulated systems, finite size of the
DMPC bilayer system and insufficient sampling. However, the fact that we have
obtained similar long time diffusion coefficient for all lipid atoms indicates
that our MD trajectory is sufficiently long to capture the linear diffusion
regime.
 \begin{table}
 \caption{\label{Dtable}The diffusion coefficient calculated from
 a linear fit to the the mean square displacement and the memory function approach
 described by Eqs.~\ref{msd-proj} and \ref{memory-function} 
 for the atoms
 shown in Fig.~\ref{lipid} and the center of mass (CM). The
 experimental value obtain from FRAP experiments is 0.69$\times 10^{-7}$ cm$^2$ s$^{-1}$ \cite{Almeida1995}.}
 \begin{ruledtabular}
 \begin{tabular}{ccc}
 atom & $D_L$-msd & $D_L$-theory\\
 \hline
 CT & 1.67$\times 10^{-7}$ cm$^2$ s$^{-1}$ & 1.25$\times 10^{-7}$ cm$^2$ s$^{-1}$ \\
 CH & 1.27$\times 10^{-7}$ cm$^2$ s$^{-1}$ & 1.15$\times 10^{-7}$ cm$^2$ s$^{-1}$\\
 P & 1.30$\times 10^{-7}$ cm$^2$ s$^{-1}$  & 1.07$\times 10^{-7}$ cm$^2$ s$^{-1}$\\
 CM & 1.46$\times 10^{-7}$ cm$^2$ s$^{-1}$ & 1.06$\times 10^{-7}$ cm$^2$ s$^{-1}$
 \end{tabular}
 \end{ruledtabular}
 \end{table}

 We propose an approach based on the Zwanzig-Mori projection operator method to
 model the MSD, $\langle[\delta{r}^a(t)]^2\rangle$, for a lipid atom $a$ in terms
 of five fitting parameters.
 We start from the equation of motion for the density, $\rho_a(q,t)$,
 autocorrelation function $\phi_s^a(q,t) = \langle \rho_a(-q,0) \rho_a(q,t)
 \rangle$ of a tagged atom $a$ at wave-vector $q$ \cite{Hansen2006}
\begin{equation}
\label{self-proj}
\partial_t^2 \phi_s^a(q,t) + q^2 v_{a}^2 \phi_s^a(q,t) 
+ \int_0^t M^a(q,t-s) \partial_s \phi_s^a(q,s) \mbox{d}s = 0.
\end{equation}
Here $v_{a} = \sqrt{k_B T/m_a}$, $k_B$ is the Boltzmann's constant, $T$ is the
temperature, $m_a$ is the mass of atom $a$, and $M^a(q,t)$ is the memory
kernel. In general, MCT refers to an approximation for $M^a(q,t)$.
The equation of motion for the MSD in 2D can be obtained from 
$\left< [\delta r^a(t)]^2 \right> = 4 \lim_{q \rightarrow 0} [1-\phi_s^a(q,t)]/q^2$
which gives
\begin{equation}
\label{msd-proj}
\partial_t \langle [\delta r^a(t)]^2 \rangle + \int_0^t M^a_0(t-s) \langle [\delta
r^a(s)]^2 \rangle \mbox{d} s = 4 v_{a}^2 t,
\end{equation}
where $M^a_0(t) = \lim_{q \rightarrow 0} M_a(q,t)$. 

We propose the following ansatz for the memory function
\begin{equation}
\label{memory-function}
M^a_0(t) = \delta(t)/\tau_3^a + \frac{B_a e^{-t/\tau_1^a}}{1 + (t/\tau_2^a)^{\beta^a}} \;,
\end{equation}
where $\delta(t)$ is the Dirac delta function. The physical meaning of the fitting
parameters $\tau_3^a$, $B_a$, $\tau_1^a$, $\tau_2^a$ and $\beta^a$ are clarified
next.

\begin{figure}
  \centering
  \includegraphics[width=3.2in]{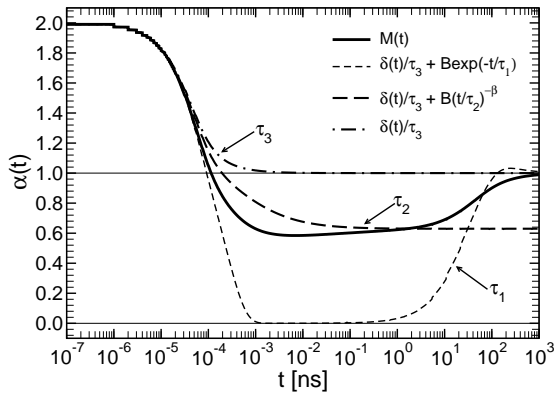}
  \caption{$\alpha(t)$ for MSDs calculated from Eq.~\eqref{msd-proj} for
    different memory kernels $M(t)$ as described in the text.}
  \label{fig:alpha}
\end{figure}

If $B = 0$, then Eqs.~\eqref{msd-proj}-\eqref{memory-function} describe the free
diffusion of a Brownian particle with a diffusion coefficient $D_{3}=k_B T/(\tau_3
m)$. Indeed, when $B=0$, Eq.~\eqref{msd-proj} can be solved analytically and
$\langle[\delta{r}(t)]^2\rangle =4D_3 t - 4D_3\tau_3 (1-e^{-t/\tau_{3}})$. For
$t\ll\tau_{3}$, $\langle[\delta{r}(t)]^2\rangle \approx (2 k_B T /m) t^2$
corresponding to the ballistic regime, while for $t\gg\tau_3$
$\langle[\delta{r}(t)]^2 \rangle\approx 4D_3 t$ corresponding to the Fickian
diffusion regime. To illustrate the crossover between these two limiting cases we
plotted $\alpha(t)$ defined by Eq.~(\ref{eq:alpha}), Fig.~\ref{fig:alpha}.  For
free diffusion $\alpha(t)$ changes smoothly from $\alpha=2$ (ballistic dynamics)
to $\alpha=1$ (Fickian diffusion) as shown in Fig.~\ref{fig:alpha} (dot-dashed
line) for $\tau_{3}=10$~fs.

By neglecting the power-law term in the memory kernel and setting
$M^0(t)=\delta(t)/\tau_3 + B \exp(-t/\tau_1)$, $\left< [\delta r(t)]^2 \right>$
can be calculated analytically, though the expression is too cumbersome to be
shown here. In this case, first the MSD crosses over from the ballistic regime
(for $t\ll \tau_3$) to a caging region (for $\tau_3 \ll t \ll \tau_1$)
characterized by a plateau in $\langle[\delta{r}(t)]^2\rangle$ (with $\alpha = 0$)
as shown in Fig.~\ref{fig:alpha} (dashed curve). Then there is a crossover from
the caging to the Fickian regimes that takes place around $\tau_1$. The extent of
the caging region is determined by the difference between the two time scales
$\tau_3$ and $\tau_1$. 

The power-law term in the memory kernel, Eq.~\eqref{memory-function}, is
responsible for the anomalous subdiffusion in the MSD. To illustrate we have
numerically calculated the MSD and the corresponding $\alpha(t)$ for $M^0(t)=
\delta(t)/\tau_3 + B/[1 + (t/\tau_2)^{-\beta}]$. As shown in Fig.~\ref{fig:alpha}
(long-dashed curve), in a broad time interval centered around $\tau_2=0.1$~ps and
extending over four decades, the dynamics slowly crosses over from ballistic to
sub-diffusive (with $\beta < 1$) behavior.

To capture all the features of the mean square displacement of lipid atoms in a
bilayer membrane one needs to consider the full kernel (\ref{memory-function})
with all three time scales.  A representative result for $\alpha(t)$ obtained with
the proposed memory function method is shown in Fig.~\ref{fig:alpha} (solid line).

\begin{figure*}
\includegraphics[width=3.2in]{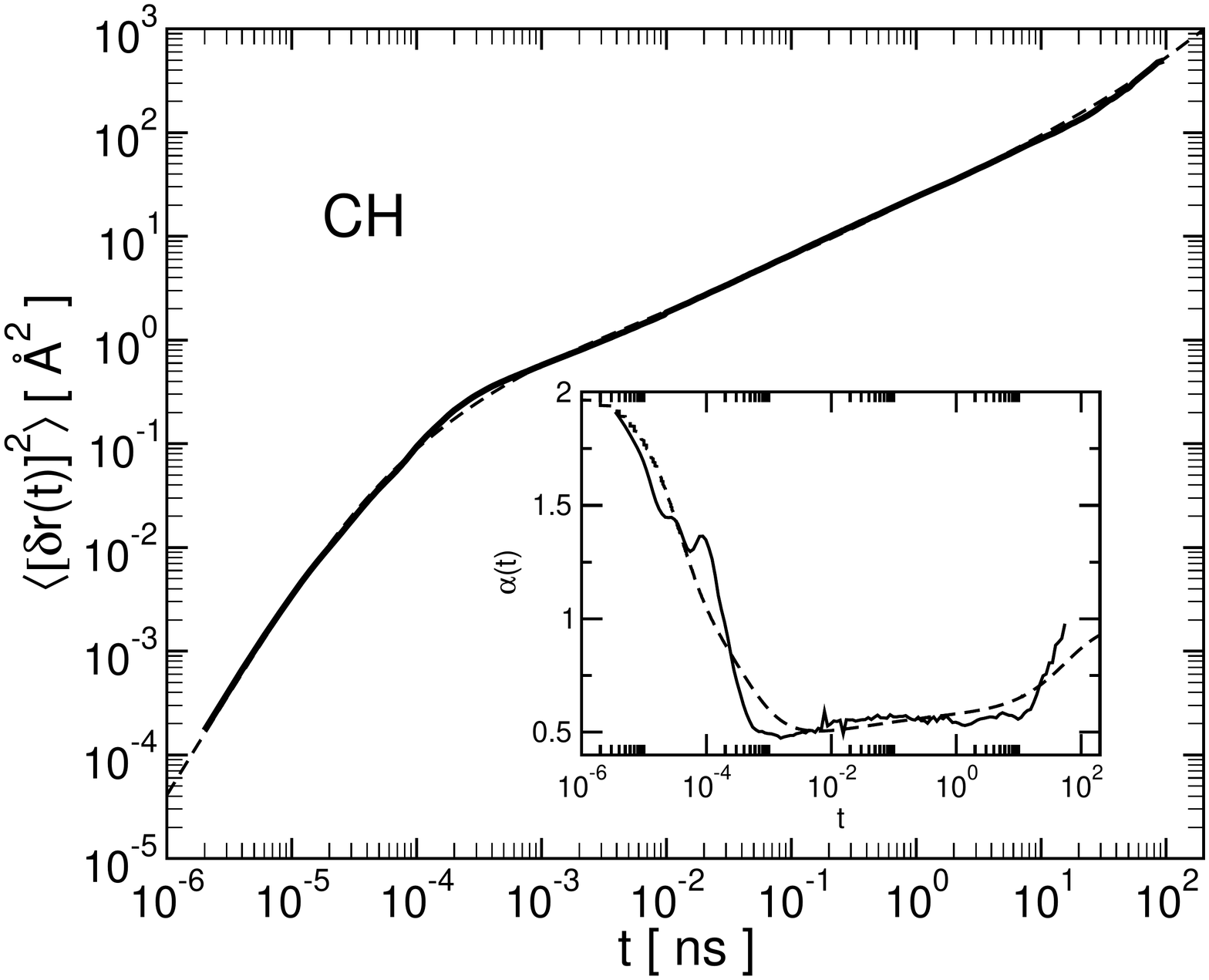}
\includegraphics[width=3.2in]{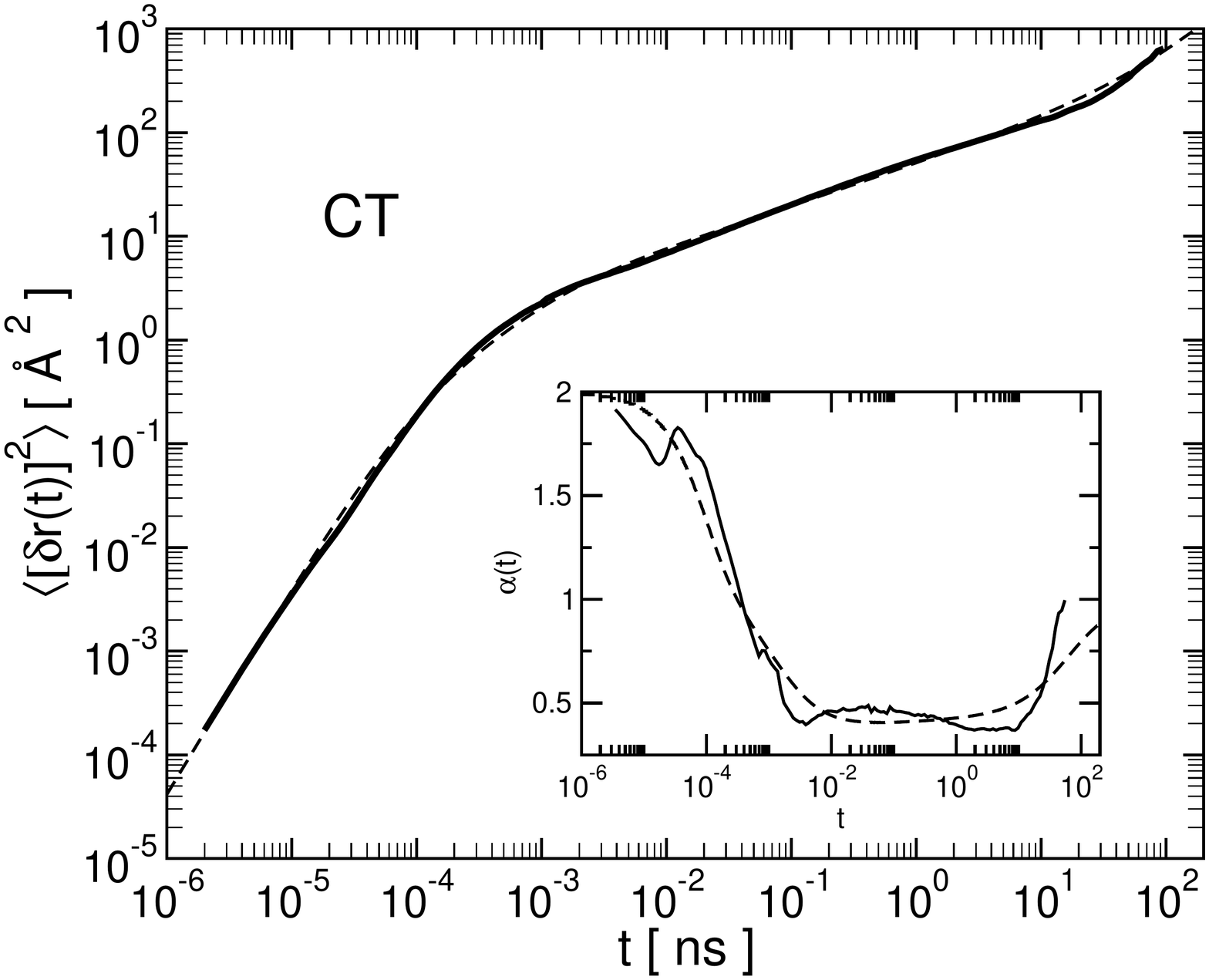}
\includegraphics[width=3.2in]{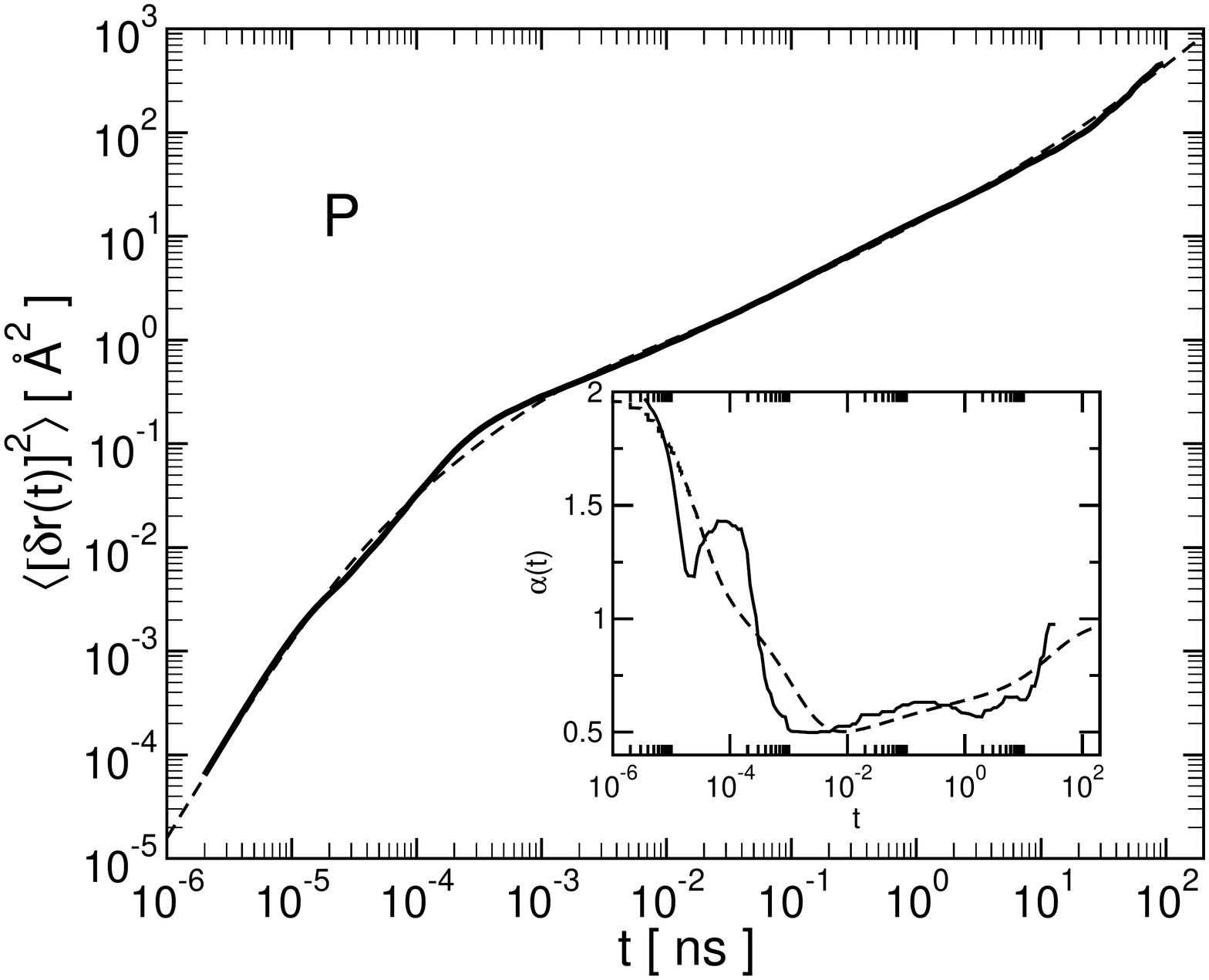}
\includegraphics[width=3.2in]{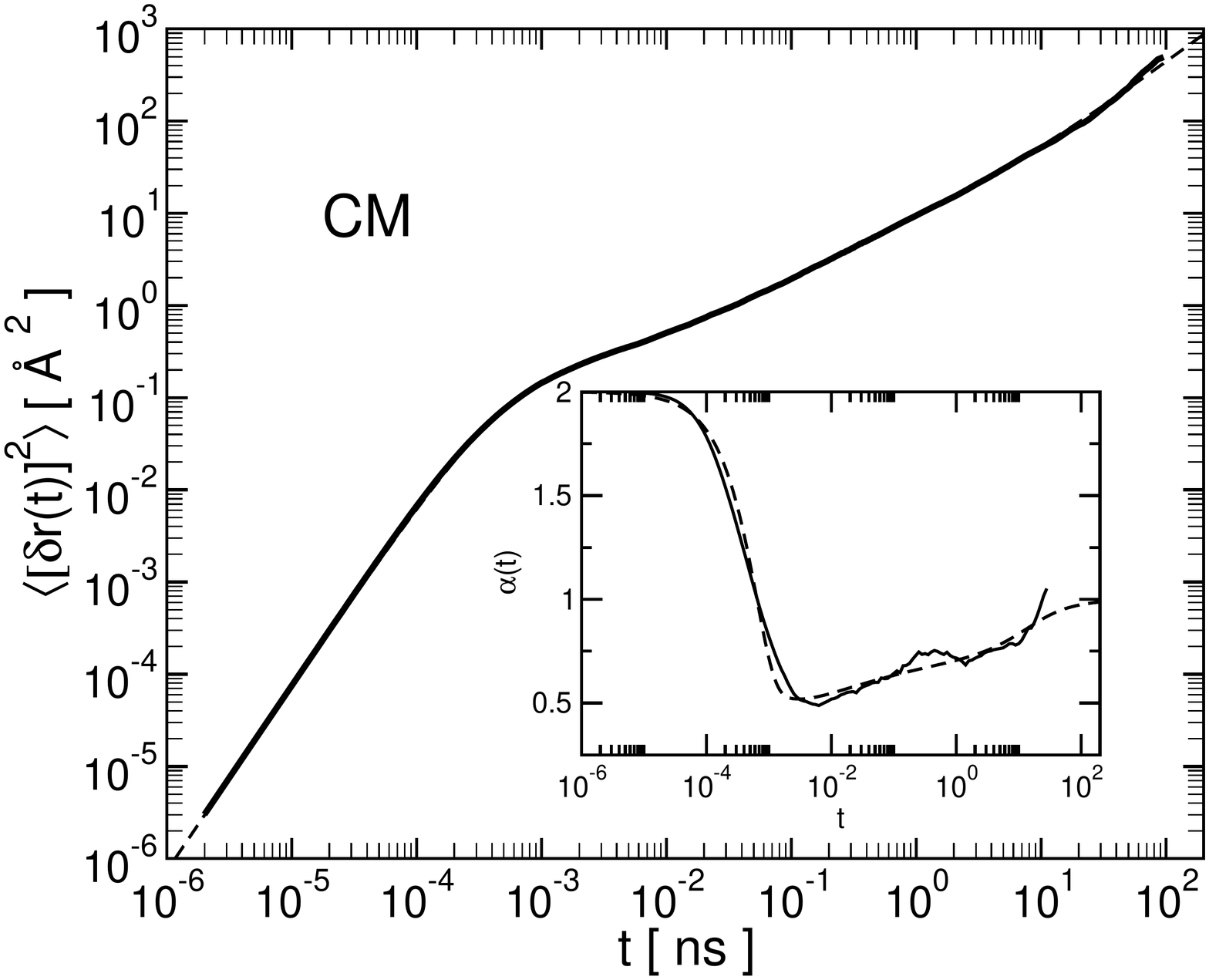}
\caption{\label{msd-alpha}The fits to the mean square displacement
using the memory function approach, Eqs.~\ref{msd-proj} and \ref{memory-function}.
The solid line is $\left< \delta r^2(t) \right>$ calculated from the simulations
and the dashed lines are the fits. The insets show the exponent 
$\alpha(t) = \partial \ln \langle [\delta r^a (t)]^2 \rangle /\partial\ln(t)$ as a 
function of time. A running average was used to decrease the noise in 
the simulation results.}
\end{figure*}

Based on the above analysis, one can identify $\tau_1$, $\tau_2$, and $\tau_3$ as
the characteristic time scales for the crossover from the subdiffusion to Fickian
diffusion region, the onset of the subdiffusion regime, and the crossover from
ballistic to caging region, respectively.
To determine the times $\tau_1$, $\tau_2$, and $\tau_3$ we performed a least
squares fit using Eqs.~\eqref{msd-proj} and \eqref{memory-function} to $\left<
  [\delta r^a(t)]^2 \right>$. The fits are shown in Fig.~\ref{msd-alpha} and the
fit parameters are given in Table \ref{msd-fit-param}.  The inset to the figures
show the corresponding $\alpha^a(t)$ for the fits and calculated from the MD
simulation. For all the individual atoms there are features in $\alpha^a(t)$ for
$t < 1$ ps that are not captured by our memory function method. The crossover to
Fickian diffusion also appears to be sharper in the simulations than predicted by
the theory.  However, the memory function theory fits the MSD from MD simulation
well for over eight decades in time.

The three time scales are separated by at least an order of magnitude, and define
three clearly separated dynamical regimes of the lipid bilayer system.
The short time scale $\tau_3 \sim 10~\text{fs}$ characterizes the crossover
between the ballistic and caging regions. The intermediate time scale $\tau_2 \sim
1~\text{ps}$ defines the onset of the sub-diffusive region. Finally, the long time
scale $\tau_1 \sim 10~\text{ns}$ identifies the crossover from subdiffusion to
Fickian diffusion.

\begin{table}
\caption{\label{msd-fit-param}Fit parameters of the mean square displacement 
using the memory function approach, Eqs.~\eqref{msd-proj} and \eqref{memory-function}.}
\begin{ruledtabular}
\begin{tabular}{cccccc}
Atom  & B (ps$^{-2}$) & $\tau_1$ (ns) & $\tau_2$ (ps) & $\tau_3$ (fs) & $\alpha$ \\
\hline
P & 83.2 & 19.99 & 2.16 & 5.62 & 0.70 \\
CH & 217 & 38.26 & 0.411 & 6.64 & 0.60 \\
CT & 42.9 & 47.02 & 0.335 & 18.1 & 0.44 \\
CM & 10.4 & 9.38 & 1.14 & 91.4 & 0.75 
\end{tabular}
\end{ruledtabular}
\end{table}

We conclude this section by noting that the memory function approach provides
another means to calculate the diffusion coefficient $D_L$.  
Inserting $\langle[\delta{r}(t)]^2\rangle\approx 4Dt$ (valid in the
$t\rightarrow\infty$ Fickian diffusion regime) into Eq.~\eqref{msd-proj}, one
finds that

\begin{equation}
D_L = v_a^2\left[\int_0^\infty M^a_0(t) dt\right]^{-1} \;.
\label{eq:D-mf}
\end{equation} 
The diffusion coefficient $D_L$ should be independent of the atom used in the
calculations.  The diffusion coefficients ($D_L$-theory) for the P, CH, CT atoms and the center
of mass calculated using Eq.~(\ref{eq:D-mf}) are given in Table
\ref{Dtable}. These values are in good agreement with those determined directly
from the MD simulation, but are systematically smaller.  Furthermore, like the
fits to the asymptotic behavior of the MSD, they are nearly equal.  Note that the
fitting parameters differ by an order of magnitude for different atoms in the
lipid, but the long time diffusion coefficient is the same for all the atoms.

\section{Incoherent Intermediate Scattering Functions}
\label{sf}

To understand the implications of the large sub-diffusive region in the analysis
of neutron scattering experiments, we examined the self (incoherent) intermediate
scattering function (SISF) for different atoms in the lipid and the lipid center
of mass. We start this section with some theoretical background that relates the
moments of atomic displacements with the SISF in terms of a cumulant expansion.
Next, we examine the viability of the second (Gaussian) and fourth order cumulant
approximations for calculating the SISF for the selected atoms studied in
Sec.~\ref{msd}. This will allow us to quantitatively relate the MSD to the SISF
and the interpretation of neutron scattering data. Finally, we perform the same
analysis for the lipid hydrogens, which dominate the incoherent neutron signal.


\subsection{Background and Theory}

The differential cross section of the quasi-elastic scattering of neutrons in a
solid angle $\mbox{d}\Omega$ with an energy transfer $\hbar \omega$ can be
expressed as \cite{Lovesey1984}
\begin{equation}
\frac{\mbox{d} \sigma^2}{\mbox{d} \Omega \mbox{d} \omega} \propto
\sum_{n} \left( b_{inc}^n \right)^2 S_s^{n}(q,\omega) +
\sum_{n, m} b_{coh}^n b_{coh}^m S_{coh}^{n m}(q,\omega),
\label{crosssection}
\end{equation}
where $n, m$ are atom-type indices, while $b_{inc}^n$ [$b_{coh}^n$] and
$S_s^n(q,\omega)$ [$S_{coh}^{n m}(q,\omega)$] are, respectively, the 
incoherent [coherent] scattering length and dynamic structure factor.
$S_s^n(q,\omega)$ contains information about the single particle motion of nuclei of
type $n$, and in principle it can be used to determine the self
diffusion coefficient $D_L$.

In a computer simulation, $S_s^n(q,\omega)$ is obtained from the Fourier transform
of the SISF
\begin{equation}
I_s^n(q,t) = \frac{1}{N_n}
\left\langle \sum_{m=1}^{N_n} e^{i \vec{q} \cdot [ \vec{r}\,_m^n(0) -
  \vec{r}\,_m^n(t)]} \right\rangle , 
\label{incint}
\end{equation}
where the summation goes over all $N_n$ atoms of type $n$ in the system.
Note that the only difference between $I_s^n(q,t)$ and $\phi_s^a(q,t)$, discussed
in Sec.~\ref{msd}, is that $I_s^n(q,t)$ is defined for a subset of atoms 
with the same scattering length, while $\phi_s^a(q,t)$ is defined for a single tagged atom.
The fourth order cumulant expansion of Eq.~\eqref{incint} reads
\begin{align}
\label{sisf-expand}
I_s^n(q,t) \approx &
    \exp \left( {-q^2 \frac{\langle [\delta{r}^n(t)]^2 \rangle}{4} } \right) \times \nonumber \\ 
   & \left[ 1 + \frac{1}{2} \left( \frac{q^2 \left< [\delta r^n(t)]^2 \right>}{4} \right)^2 \gamma_2^n(t) \right],
\end{align}
where 
\begin{equation}
\label{nongauss}
\gamma_2^n(t) = \frac{\left< [\delta r^n(t) ]^4 \right>}{2 \left< [\delta r^n (t)]^2 \right>} - 1.
\end{equation}
Equations \eqref{sisf-expand} and \eqref{nongauss} express the SISF is terms of the MSD and
the non-Gaussian parameter $\gamma_2^n(t)$.
If the distribution of lateral displacements of the lipid atoms $G_s^{n}(r,t) =
\langle \delta(r-|\vec{r}^{\,n}(t)-\vec{r}^{\,n}(0)|) \rangle$ is Gaussian in
space, then
\begin{equation}
I_s^n(q,t) = \exp \left( -q^2\frac{\langle[\delta r^n(t)]^2 \rangle}{4} \right).
\label{selfgauss}
\end{equation}
For 2D Fickian diffusion $\langle[\delta{r}^n(t)]^2 \rangle = 4D_L t$, $G_s^n(r,t)$
is Gaussian in space, and
\begin{equation}
  \label{selfdiff}
  I_s^n(q,t) = e^{-q^2 D_L t} \;.
\end{equation}
Note that, while the distribution function of the displacements of atoms that
undergo Fickian diffusion is Gaussian (having a width that increases as
the square root of time), the converse in not true in general.

The dynamic structure factor corresponding to the SISF given by
Eq.~\eqref{selfdiff} is a Lorentzian \cite{Hansen2006}
\begin{equation}
S_s^n(q,\omega) =\frac{1}{\pi} \frac{q^2 D_L}{(q^2 D_L)^2 + \omega^2} \;,
\label{incD}
\end{equation}
which is the basis of evaluating the diffusion coefficient
$D_L$ from neutron scattering experiments.
To determine the diffusion coefficient, it is generally assumed that the
translational motion of the lipid is decoupled from other motions of the atoms
within the lipid (e.g., vibrations, rotations, conformational changes, etc.). The
resulting $S_s^n(q,\omega)$ is a convolution of these other motions with
Eq.~\eqref{incD}.
From the above discussion, however, it should be clear that the reliable
experimental determination of the diffusion coefficient $D_L$ based on
Eq.~\eqref{incD} is possible only when the MSD of the lipid atoms and/or the lipid
center of mass increases linearly in time.

\begin{figure}
\includegraphics[width=3.0in]{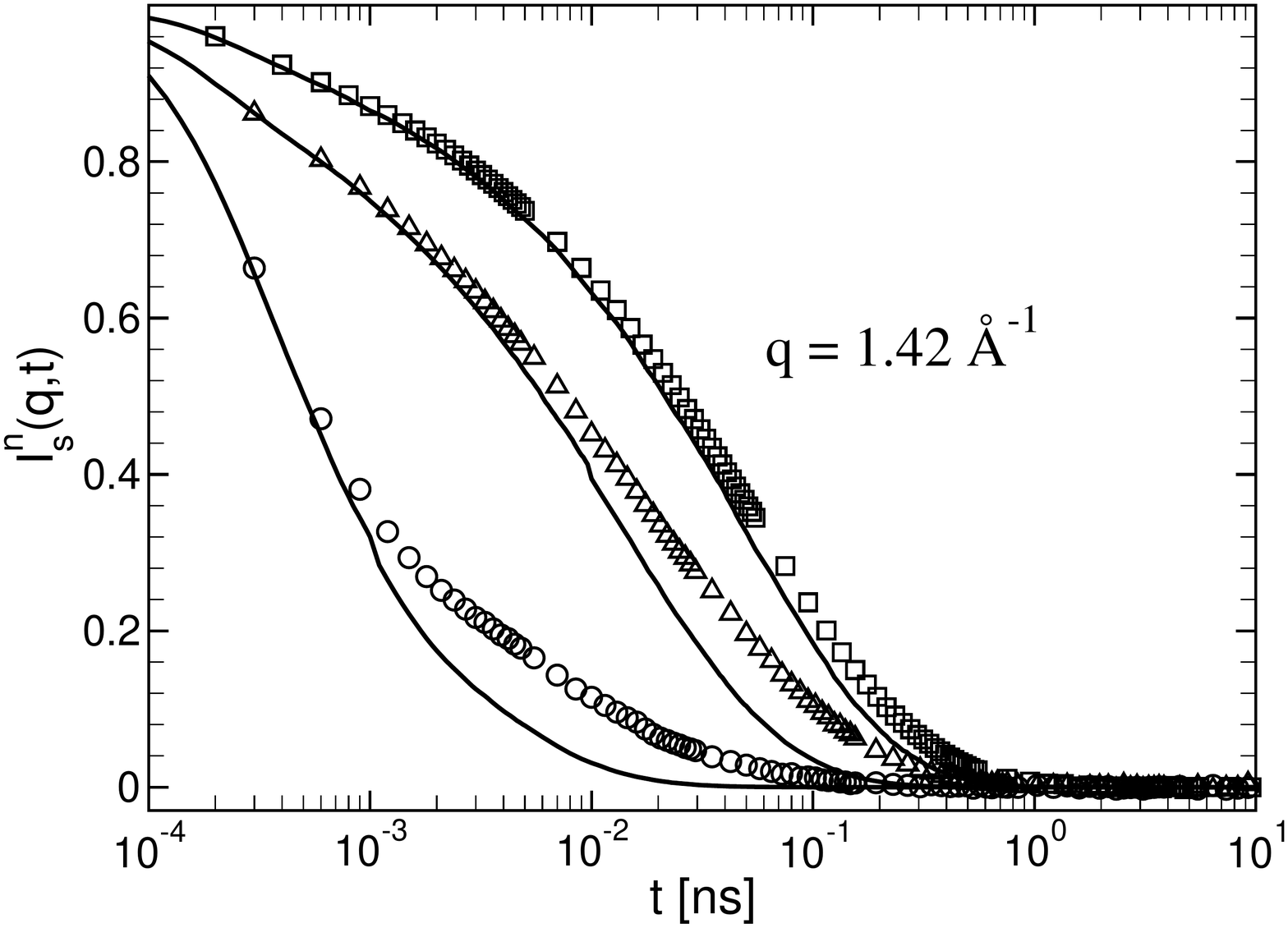}
\includegraphics[width=3.0in]{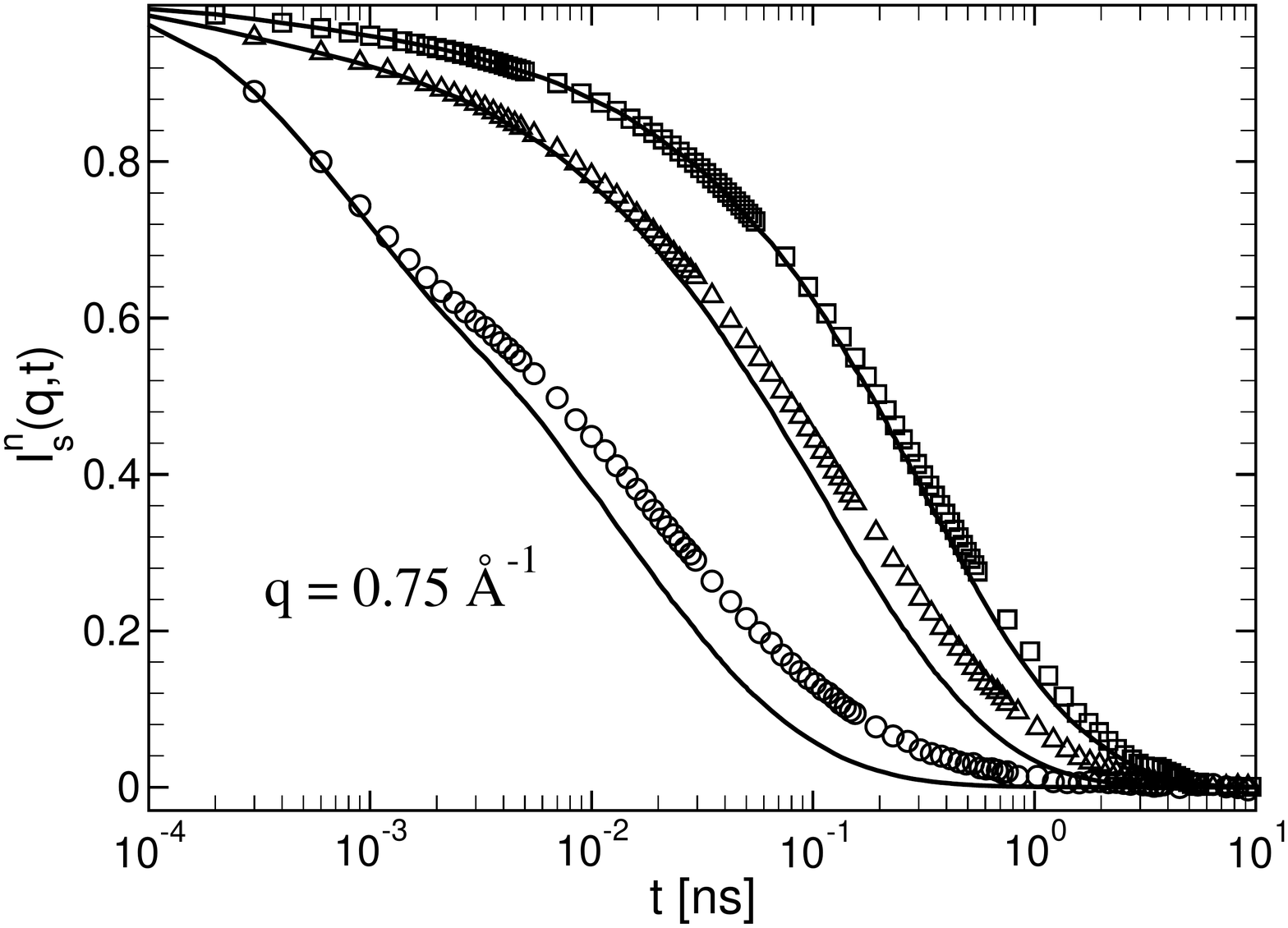}
\includegraphics[width=3.0in]{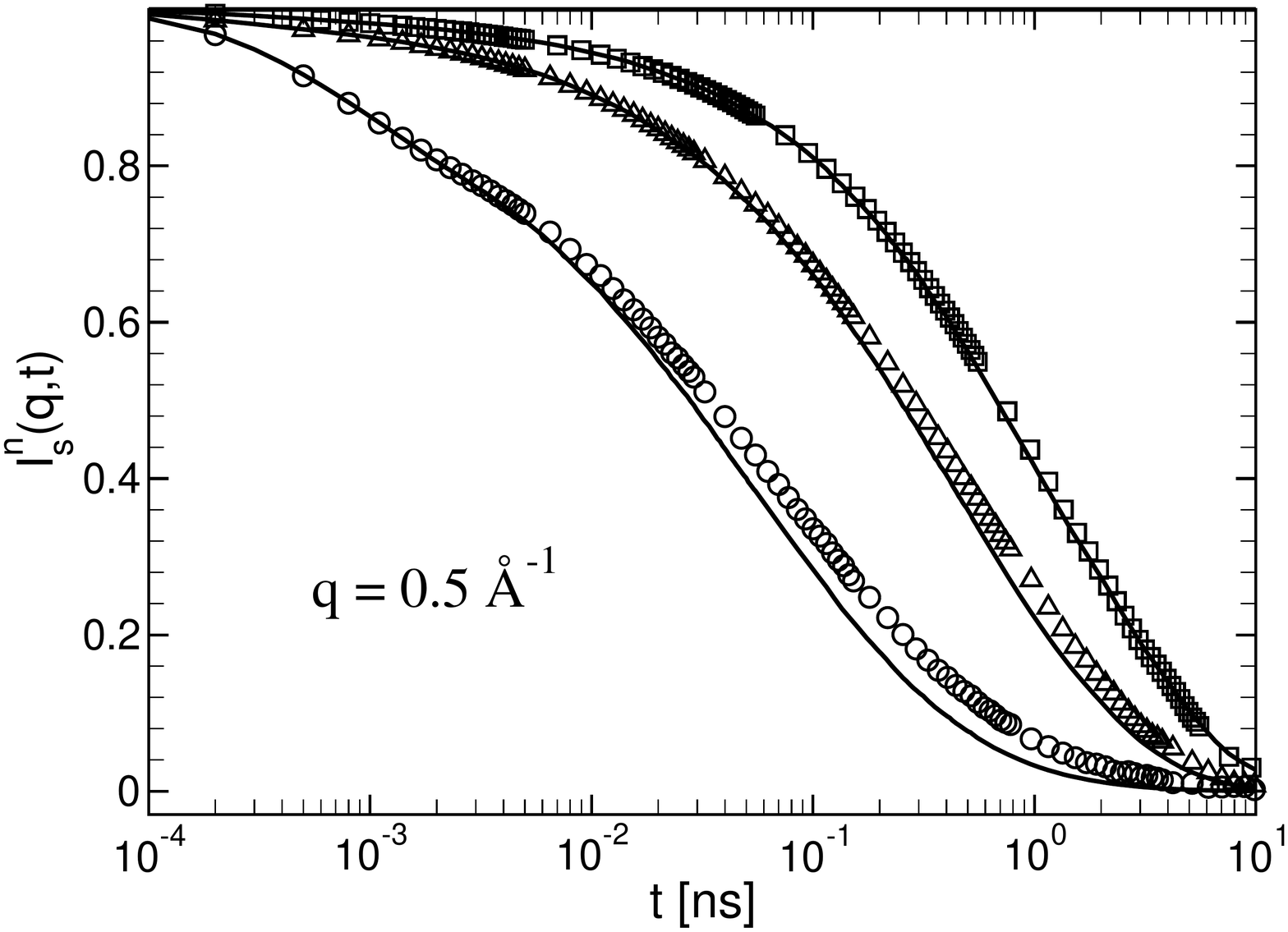}
\caption{Self intermediate scattering function for P (squares), CH (triangles) and
  CT (open circles) atoms for $q=1.42$\AA$^{-1}$ (top), $q=0.75$\AA$^{-1}$
  (middle) and $q=0.5$\AA$^{-1}$ (bottom). The solid curves correspond to the
  Gaussian approximation Eq.~\eqref{selfgauss}.}
\label{inc}
\end{figure}

\subsection{Individual Atoms}

In this sub-section we calculate the SISF for the P, CH and CT atoms in the lipid,
and examine the validity of the expansion given by Eq.~\eqref{sisf-expand} for
these atoms. This analysis will give insight into the different dynamics along the
lipid and aid in the analysis of the dynamics of the hydrogen atoms discussed in
the next sub-section.

Shown as symbols in Fig.~\ref{inc} is $I_s^n(q,t)$ for $q\in\{1.42, 0.75,
0.5\}$~\AA$^{-1}$ calculated for the selected P, CH and CT atoms. The peak in the
static structure factor for the lipid tails (head groups) occurs around
$q=1.42$\AA$^{-1}$ ($q=0.75$\AA$^{-1}$).  The Gaussian approximation (solid
lines), Eq.~\eqref{selfgauss}, of the SISF appears to be excellent for the P atoms
(open squares) for all $q$ values considered, but noticeable deviations are
present for the carbon (CH and CT) atoms.  Note, however, that as $q$ decreases
the Gaussian approximation becomes increasingly better for the carbon atoms as
well. 

\begin{figure}
\includegraphics[width=3.2in]{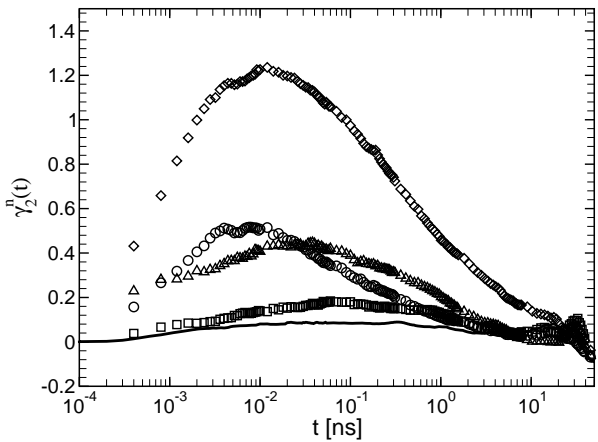}
\caption{Time dependence of the non-Gaussian parameter $\gamma_2^n(t)$ for the P
  (square), CT (circle) and CH (triangle) atoms, as well as, averaged over all the
  carbon atoms within the lipid (diamond), and calculated for the center of mass
  (solid line).}
\label{ng}
\end{figure}

In order to quantify how well the distribution function $G_s^n(r,t)$ of the
lateral displacements of lipid atoms of type $n$ can be approximated by a
Gaussian, we calculate the non-Gaussian parameter $\gamma_2^n(t)$ given in
Eq.~\eqref{nongauss}.
Shown in Fig.~\ref{ng} are $\gamma_2^n(t)$ for the atoms P, CH, CT, for all the
carbon atoms within the lipid, and for the center of mass of the lipids. For the
carbon atoms, there is a peak in $\gamma_2^n(t)$ around the beginning of the
sub-diffusive region in the MSD, (i.e., $t\approx 10$~ps). The peak heights for CH
and CT atoms are at least a factor of three smaller than when the displacements
are averaged over all the carbon atoms.  For times when $\gamma_2^n(t)$ is small,
Eq.~\eqref{selfgauss} is a good approximation, which can be seen by comparing
Figs.~\ref{inc} and \ref{ng}.  Furthermore, it is evident that the approximation
given by Eq.~\eqref{selfgauss} is very poor when applied to the average of the
displacements of all the carbon atoms within the lipid.  Notice that
$\gamma_2^n(t) \approx 0$ when $t \approx 30$ ns for all $n$, which is the time
scale for the crossover to the linear region in the MSD ($\tau_1$ in
Table~\ref{msd-fit-param}) found in Sec.~\ref{msd}.

\begin{figure}
  \centering
  \includegraphics[width=3.2in]{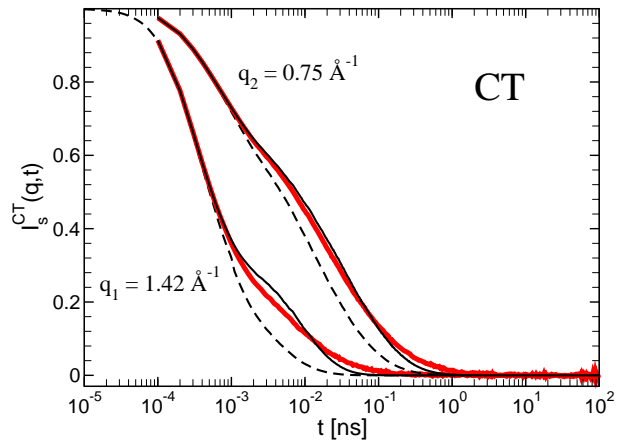}
  \caption{(Color online) SISF for the CT atoms for two
    scattering vectors $q_1=1.42$ \AA$^{-1}$ and $q_2=0.75$ \AA$^{-1}$. 
    The dashed curves represent the Gaussian approximation
    Eq.~\eqref{selfgauss}, the thin-solid curves the non-Gaussian approximation
    Eq.~\eqref{sisf-expand}, and the thick (red) curves are the exact SISF
    calculated using Eq.~\eqref{incint}.}
  \label{fig:Is-ng}
\end{figure}

We examined the approximation given by Eq.~\eqref{sisf-expand} for CT, the atom
with the largest value of $\gamma_2^n(t)$.
Note that for short and long times $G_s^n(r,t)$ is approximately Gaussian is
space, which is evident by the small values of the non-Gaussian parameter
$\gamma_2^{CT}(t)$ for these times (see Fig.~\ref{ng}). For intermediate times
that correspond to the subdiffusion dynamics, $\gamma_2^{CT}(t)$ is finite
(especially for the carbon atoms) and, according to Eq.~(\ref{sisf-expand}), one
expects noticeable deviation from the Gaussian approximation,
Eq.~\eqref{selfgauss}.
A comparison between the direct calculation, the Gaussian approximation given by
Eq.~\eqref{selfgauss}, and the first non-Gaussian correction given by
Eq.~\eqref{sisf-expand} of the SISF for the CT atoms is shown in
Fig.~\ref{fig:Is-ng} for two scattering vectors $q_1=1.42$ \AA$^{-1}$ and
$q_2=0.75$ \AA$^{-1}$. The figure shows that the first correction term matches
very well the direct calculation of $I_s^n(q,t)$ for all $t$ and both $q$ values,
while the Gaussian approximation is rather poor for intermediate times.

Because of the relationship between the MSD and the SISF given by
Eq.~\ref{sisf-expand}, one may conclude that the stretched exponential form of
$I_s^n(q,t)$ is an intrinsic property originating from the polymer nature of the
lipid molecules.
For $t$ values when $G_s^n(r,t)$ is close to a Gaussian, the SISF for lipid atoms
can be well approximated by Eq.~\eqref{selfgauss}. This agreement, combined with
the behavior of the MSD from Sec.~\ref{msd} and the small values of
$\gamma_2^n(t)$ for all $t$, suggests that the decay of the SISF for the
phosphorus (P) atoms should be well described by a Kohlraush-William-Watt (KWW)
function $A(q) \exp[-[t/\tau(q)]^{\beta(q)}]$.  The parameters $\beta(q)$ and
$\tau(q)$ have a nontrivial $q$ dependence. Larger $q$ corresponds to smaller
displacements and, hence, a shorter time scale. Smaller $q$ probes displacements
over larger length scales, thus it corresponds to dynamics on longer time
scales. From Eq.~\eqref{selfgauss} and the fits to $\langle [\delta r^P(t)]^2
\rangle$ from Section \ref{msd}, we expect that $\beta(q) \approx 0.6$ for a range
of $q$ values, and we determine this range later in this section. For $t\gtrsim
30~\text{ns}\sim\tau_1$ the SISF crosses over to the simple exponential form
characterized by $\beta(q)=1$.
However, this Fickian diffusion regime can be observed only for small $q$ values,
which computationally are quite expensive to reach due to the long MD simulation
time required to calculate the decay of the SISF. Thus, in most all-atom MD
simulations the mainly explored dynamic region is the sub-diffusive one.

\begin{figure}
\includegraphics[width=3.2in]{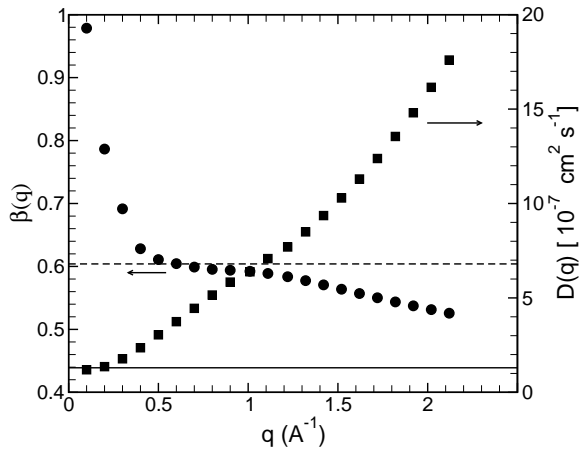}
\caption{The $q$ dependence of $\beta(q)$ (circles and left axis) and $D(q) =
  1/[\tau(q) q^2]$ (squares and right axis) for the P atoms from fits of
  $I_s^P(q,t)$ to $\exp[-[t/\tau(q)]^{\beta(q)}]$. The dashed line is 
  $\beta = 0.6$ which was obtained from the MSD. The solid line is the 
  diffusion coefficient $D_L$-msd calculated from the linear region of the 
  MSD for the phosphorus atom.}
\label{betaD}
\end{figure}

To examine the $q$ dependence of $\beta(q)$, and $\tau(q)$ we fit the SISF
calculated for the P atoms for $t > 10~\text{ps}$ to $A(q)
\exp[-[t/\tau(q)]^{\beta(q)}]$.  Shown in Fig.~\ref{betaD} is $\beta(q)$ (circles
and left axis).
For $q = 0.6$ \AA$^{-1}$, we
found $\beta(q)\approx 0.6$ (dashed line), which is in agreement with the exponent 
calculated from Eq.~\eqref{eq:alpha} (i.e., the 
sub-diffusive exponent of the MSD). 
For smaller $q$ values, $\beta(q)$ increases sharply
towards the value $\beta=1$.
Note that the diffraction peak for the lipid head groups has been observed to be
around $q \approx 0.75$ \AA$^{-1}$, which corresponds to the lipids being separated by
approximately $9$ \AA. However, the stretched exponential fits yield
$\beta(q) = 1$ only for $q \lesssim 0.1$ \AA$^{-1}$, which corresponds to a length scale of
$62$ \AA, around nine DMPC lipid diameters. Notice that $\beta(q)$ is approximately equal
to the exponent for the sub-diffusive dynamics of the lipids for $q$ values
around the diffraction peak for the lipid head group.

In the $q\rightarrow 0$ limit, $D(q) = 1/[\tau(q) q^2]$ is equal to the lateral
self-diffusion coefficient $D_L$ obtained from the asymptotic behavior of the MSD.
$D(q)$ is also shown in Fig.~\ref{betaD} (squares and right axis) from the results
of the fits to $I_s(q,t)$ for the P atoms.
Shown as a solid line in Fig~\ref{betaD} is the value of the diffusion coefficient
$D_L^{P}$-msd given in Table~\ref{Dtable} for the $P$ atoms.  It should be noted
that, after a steep decrease with $q$, $D(q)$ in Fig~\ref{betaD} seems to level
off for $q\lesssim 0.2$ \AA$^{-1}$ and is close to the value obtained from the
linear fit to the MSD for the Phosphorus atom and the center of mass.

\subsection{Lipid Hydrogen Atoms}

According to Eq.~\eqref{crosssection} the quasi-elastic scattering signal is
modulated by the scattering cross-sections $b_{inc}$ and $b_{coh}$. To determine
$D_L$ one has to measure the incoherent dynamic structure factor, which is dominated
by the hydrogen atoms in the lipids. Therefore, we examine $I_s^H(q,t)$ calculated
for all the hydrogen atoms in the lipids.

The light hydrogen (H) atoms follow the motion of the heavy atoms. However, being
covalently bound to the heavy atoms in the lipid, the H-atoms also perform vibrations
and rotations about these atoms. Thus, one expects that $I_s^H(q,t)$
deviates (especially at shorter times) noticeably from $I_s^n(q,t)$ for the
carbon, phosphorus, or oxygen atoms. Nevertheless, $I_s^H(q,t)$ should have
similar characteristics in that stretched exponential relaxation
is expected for $0.4 \lesssim q \lesssim 2.5$ and there should be a crossover to
exponential relaxation for small enough $q$. 
Furthermore, we found that on ps time scales the dynamics of the lipid atoms
depend on the location of the atom in the lipid . Thus, the motion of the
hydrogen atoms will also depend on the location of the atom within the lipid in
the ps time scale. In this section we examine what can be learned from the
SISF about the dynamics of the hydrogens and the lipids.

\begin{figure}
\includegraphics[width=3.2in]{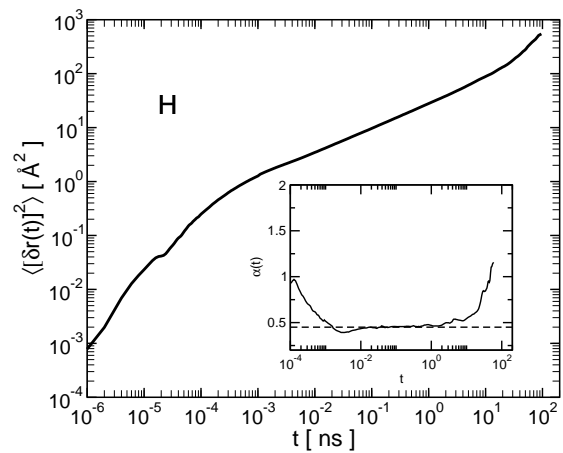}
\caption{The MSD calculated using all the hydrogen atoms in the lipids. The inset
is $\alpha(t)$ calculated using Eq.~\eqref{eq:alpha} and the dashed 
line is at the calculated value of $\beta = 0.45$.\label{Hmsd}}
\end{figure}

Shown in Fig.~\ref{Hmsd} is the MSD calculated using all the lipid hydrogens.  The
short time ballistic, the sub-diffusive region, and the diffusive region are
clearly seen. Also shown in the figure is $\alpha^H(t)$ given by
Eq.~\eqref{eq:alpha}, and we find $\beta^H = 0.45$ (dashed line in the inset). It
is important to realize that this value of $\beta$ is much different than for the
center of mass.

\begin{figure}
  \centering
  \includegraphics[width=3.2in]{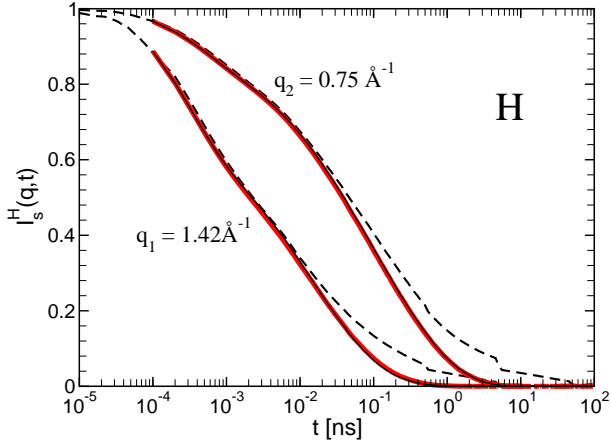}
  \caption{(Color online) SISF for the lipid hydrogen H atoms for two
    scattering vectors $q_1=1.42$\AA$^{-1}$ and $q_2=0.75$\AA$^{-1}$.
    The dashed curves represent the Gaussian approximation
    Eq.~\eqref{selfgauss}, the thin-solid curves the non-Gaussian approximation
    Eq.~\eqref{sisf-expand}, and the thick (red) curves are the exact SISF
    calculated using Eq.~\eqref{incint}.}
  \label{fig:Is-ng-H}
\end{figure}

Shown in Fig.~\ref{fig:Is-ng-H} is the direct calculation of $I_s^H(q,t)$ (solid
line), the Gaussian approximation given by Eq.~\eqref{selfgauss} (dashed line),
and the expansion given by Eq.~\eqref{sisf-expand} (red line) for wave-vectors
around the diffraction peak for the lipid tails ($q=1.42$ \AA$^{-1}$) and for the
lipid head group ($q=0.75$ \AA$^{-1}$). For the H-atoms the Gaussian
approximation, Eq.~\eqref{selfgauss}, is rather poor in the $\tau_2\sim
0.01~\text{ns}$ to $\tau_1\sim 10~\text{ns}$ range, but the first non-Gaussian
correction, Eq.~\eqref{sisf-expand}, is a very good approximation for all times.

\begin{figure}
\includegraphics[width=3.2in]{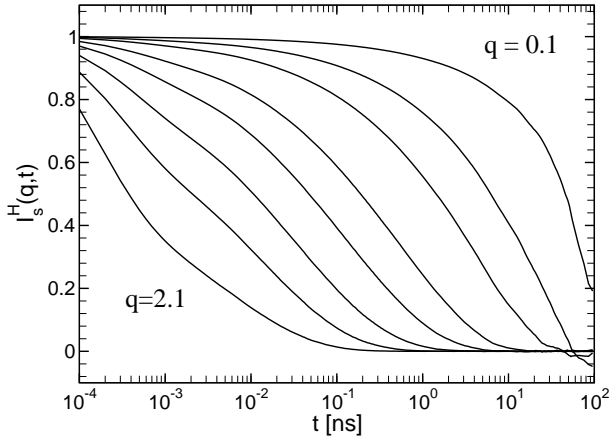}
\caption{\label{Hinc}The self intermediate scattering function calculated for the
hydrogen atoms in the lipid for $q=$ 2.1, 1.4, 1.0, 0.7, 0.5, 0.3, 0.2, and 0.1 \AA$^{-1}$
listed from left to right.}
\end{figure}

\begin{figure}
\includegraphics[width=3.2in]{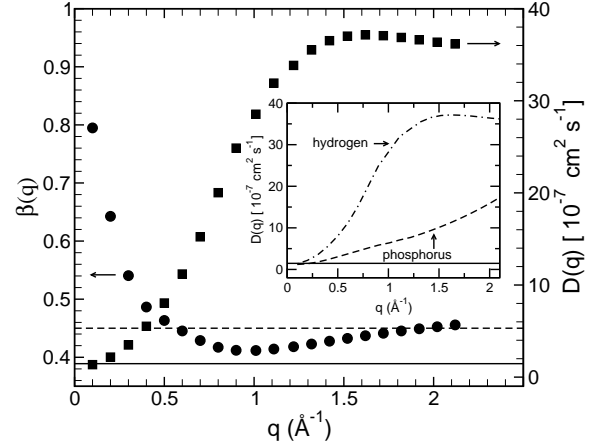}
\caption{\label{betaDh} The stretching exponent $\beta(q)$ and
$D(q)$ from fits to the SISF calculated
using the hydrogen atoms in the lipid. The dashed line is $\beta = 0.45$
which was determined from the MSD, see Fig.~\ref{Hmsd}. 
The solid line is $D_L$-msd for the center of mass. The inset is a comparison
of $D(q)$ for the phosphorus atoms (dashed line) and the hydrogen atoms
(dashed dotted line).}
\end{figure}

Shown in Fig.~\ref{Hinc} is $I_s^H(q,t)$ for several values of $q$. For large $q$
there appears to be two stages to the decay of $I_s^H(q,t)$, namely an initial
faster decay that is very nearly exponential, and a slower decay that is best
described by a stretched exponential for $t > 10$~ps, i.e., 
when the sub-diffusive behavior begins in the MSD (see
Fig.~\ref{Hmsd}).  For this analysis we will focus on the slower decay.  Again, we
fit $I_s^H(q,t)$, for $t>10~\text{ps}$, to $A(q) \exp[-[t/\tau(q)]^{\beta(q)}]$
and the results for $\beta(q)$ and $D(q) = 1/[q^2 \tau (q)]$ are shown in
Fig.~\ref{betaDh}.

For $q = 1.4$ \AA$^{-1}$ (the peak value for the static structure factor for
carbon atoms in the lipid tails) we found $\beta(q) \approx 0.43$, which is close
to but lower than the value of $\beta = 0.45$ obtained from the MSD through
Eq.~\eqref{eq:alpha} (dashed line).
For the distances corresponding to these $q$ values, the lipid is undergoing
sub-diffusive motion, Fig.~\ref{Pmsd}. For $q \approx 0.5$, $\beta(q)$ rapidly
increases to a value close to one, and $D(q)$ appears to be leveling off, but our
simulations are to short to accurately obtain the asymptotic value.
The solid line in the figure corresponds to $D_L^{CM}$-msd $ =1.62 \times 10^{-7}$
cm$^2$ s$^{-1}$ obtained from a linear fit to the MSD for the center of mass.
$D(q)$ increases faster with $q$ for the calculation using all the hydrogens than
for the P atoms (see inset to Fig.~\ref{betaDh}), but, as expected, the two
approach the same value for smaller $q$.  For small $q$, $D(q)$ is independent of
the atoms used in the calculation, and it is from these values of $q$ where
the diffusion coefficient should be obtained.  

By looking at $D(q)$ for the phosphorus and hydrogen atoms a clear pattern
emerges. The diffusion coefficient obtained from the relaxation time would appear
to be larger when determined from larger values of $q$ and for shorter times.
We examined the literature to determine if this observation is compatible with
experimental findings, and some indicative results are shown in Table
\ref{table:expD} for 1,2-Dipalmitoyl-sn-Glycero-3-Phosphocholine (DPPC). DMPC and
DPPC are often used as model systems. While DPPC has slightly larger acyl chains,
as compared to DMPC (16 to 14 C atoms), our results for DMPC should be comparable
to those for DPPC, showing the same trends.
By comparing experiments that differ by only the $q$ range (experiments 2 and 3) we
see that the larger $D_L$ is found for the larger $q$. If we compare experiments
that differ by basically the time scale involved (experiments 1 and 2) the
associated $D_L$ is larger for the shorter time scale. While the $D_L$ listed are
not comparable with our calculated $D(q)$, the experiments are in qualitative
agreement with our analysis.
\begin{table*}
 \caption{Diffusion coefficients for DPPC in the fluid phase determined by neutron scattering for 
 different time and length scales.\label{table:expD}}
 \begin{tabular}{c|c|c|c|c|c}
 Experiment & Technique &  T & length scale & time scale & $D$ \\
 \hline\hline
  1 \cite{Koenig1995} & QENS (IN5)  & 45-55 $^{\circ}$C & 0.2-1.6 \AA$^{-1}$ & 66 ps (63 $\mu$eV) & $3.5 \times 10^{-6}$ cm$^2$/s\\
  2 \cite{Pfeiffer1989} &  QENS (IN10)  & 41 $^{\circ}$C & 0.07-2 \AA$^{-1}$ & 5 ns (0.8 $\mu$eV) & 1.8$\times 10^{-7}$ cm$^2$/s\\
  3 \cite{Koenig1992} & QENS (IN10)  & 41 $^{\circ}$C & 1-2 \AA$^{-1}$ & 5 ns (0.8 $\mu$eV) & 1.6$\times 10^{-6}$ cm$^2$/s\\
 \end{tabular}
 \end{table*}

\section{Conclusions}
\label{sec:conc}

In spite of extensive literature dealing with the experimental and
computational study of the dynamics of lipid bilayers, the determination of the
lateral self-diffusion coefficient $D_L$ of lipid molecules in the membrane
remains a controversial problem. The results for $D_L$ appear to depend
significantly on the time and length scales probed by the employed measurement or
computational method. This observation prompted researchers to propose different
models that have been predominantly influenced by models of diffusion in dense
fluids.  However, lipid molecules are polymers, characterized by flexibility and
connectivity, and their lateral diffusion in the leaflets of lipid bilayers is
qualitatively different from the diffusion of molecules in dense fluids. This
difference is reflected in the time dependence of the MSD.
We observe a broad sub-diffusive region, not present in simple fluids, between the
short time ballistic region and the long time Fickian diffusion region.

Here we developed a phenomenological memory function approach, which can be used
to describe the behavior of the MSD of lipid atoms and molecules over the whole
time interval spanning from the ballistic regime (t $\lesssim 10~\text{fs}\sim
\tau_3$) to the Fickian diffusion regime (t $\gtrsim 30~\text{ns}\sim\tau_1$).
By fitting the mean square displacement, we were able to identify three clearly
separated time scales that correspond to three different dynamic regimes in the
lipid system.  Overall, our memory function fits to the MSD matched remarkably
well $\langle[\delta{r}(t)]^2\rangle$ from the MD simulation. However, some
features in the fairly broad crossover region between the ballistic and
sub-diffusive regimes are missed by our approximation.  Our memory function
approach allows us to calculate the lateral self-diffusion coefficient $D_L$ of
lipid atoms and molecules through Eq.~\eqref{eq:D-mf}, and shows that, while the
dynamics of the different atoms are very different at ps time scales, the motion
at $t \gtrsim 30~\text{ns}\sim\tau_1$ is best described by simple diffusion.

We also investigated the consequence of the sub-diffusive behavior of the MSD by
examining the first two terms in the cumulant expansion of the SISF. The first
term is referred to as the Gaussian approximation and is exact if the probability
distribution $G(r,t)$ of the displacements is Gaussian. By calculating the first
non-Gaussian correction, characterized by the parameter $\gamma_2^n(t)$
(Eq.~\ref{nongauss}), we found that $G(r,t)$ is nearly Gaussian outside the
sub-diffusive region. While for the P atoms and the center of mass of individual
lipid molecules $G(r,t)$ remained nearly Gaussian even in the sub-diffusive
region, $G(r,t)$ for the carbon and hydrogen atoms showed significant deviation
from Gaussian in this region. Interestingly, a Gaussian distribution for the
center of mass, whose width did not increase as the square root of time, for lipid
displacements has been observed in previous simulations
\cite{Lindahl2001,Wohlert2006}.

When $G(r,t)$ is Gaussian, the SISF (e.g., measured in inelastic neutron
scattering (INS) experiments of protonated lipid membrane samples) can be
expressed in terms of the MSD, Eq.~\eqref{selfgauss}.
The sub-diffusive dynamics strongly influences the scattering from the lipids
whenever the probing scattering vector, $q$, and frequency, $\omega$, correspond
to lengths and times in the sub-diffusive regime of the lipid atoms and
molecules. In this case the SISF can be fairly well fitted with a stretched
exponential (KWW) function, $A(q) \exp[-[t/\tau(q)]^{\beta(q)}]$. Note, however,
that for a given $q$ the stretched exponential will not be able to reproduce the
short (ballistic) and long (Fickian diffusion) time regimes.
We examined the $q$ dependence of $D(q) = 1/q^2 \tau(q)$ and $\beta(q)$. For the
phosphorus atoms we find that $\beta(q) \approx 0.6$ for $q\approx
0.6~\text{\AA}^{-1}$, which agrees with the exponent determined from the MSD.

We emphasize that in most INS experiments on lipid bilayers primarily one explores
the sub-diffusive region and not the long time Fickian diffusion region, thus
making the use of Eq.~\eqref{incD} inadequate. Therefore, it should come
as no surprise that the values of the lipid lateral self-diffusion coefficient
extracted from INS experiments using Eq.~\eqref{incD} have a strong $q$ dependence
and generally overestimate the real $D_L$ obtained by other (e.g., FRAP)
experiments.

Furthermore, it would be interesting to investigate the temperature dependence of
the analysis presented in this paper. In some liquids, it has been observed that
activated events become more pronounced as the temperature is lowered.  It would
be reasonable to expect a similar behavior in lipid membranes, thus the diffusion
could change character at lower temperatures and in the gel phase. It is also
unknown how the picture presented in this work would change in the gel phase of
the lipid membrane. Since the tails of the lipids are aligned, do the lipids
behave as if they are more rigid? This may change the sub-diffusive region
dramatically.  The addition of cholesterol or proteins to the lipids may also
change the nature of the sub-diffusive region and dramatically change the dynamics
of the lipids.

\section*{Acknowledgments}
We gratefully acknowledge the computer time that was generously provided by the
University of Missouri Bioinformatics Consortium. 
E.F. acknowledges partial support of NSF Grant No. CHE 0517709.




\end{document}